\documentclass[manuscript]{acmart}
\AtBeginDocument{%
  }

\usepackage{array}
\newcolumntype{L}[1]{>{\raggedright\let\newline\\\arraybackslash\hspace{0pt}}m{#1}}
\newcolumntype{C}[1]{>{\centering\let\newline\\\arraybackslash\hspace{0pt}}m{#1}}
\newcolumntype{R}[1]{>{\raggedleft\let\newline\\\arraybackslash\hspace{0pt}}m{#1}}
\usepackage{colortbl} 
\definecolor{mypink2}{RGB}{239, 240, 255}

\usepackage{mdframed}
\usepackage{listings}
\usepackage{multirow}
\begin{document}

\title[Exploring Humanlike Interactions During Collaborative Writing with an Artificial Agent]{"It Felt a Bit Eerie": Exploring Humanlike Interactions During Collaborative Writing with an Artificial Agent}


\author{Michael Yin}
\affiliation{
  \institution{University of British Columbia}
  \city{Vancouver}
  \state{BC}
  \country{Canada}
}
\email{jiyin@cs.ubc.ca}
\orcid{0000-0003-1164-5229}

\author{Angela Chiang}
\affiliation{
  \institution{University of British Columbia}
  \city{Vancouver}
  \state{BC}
  \country{Canada}
}
\email{achi2048@student.ubc.ca}
\orcid{0009-0008-8960-5837}

\author{Samuel Rhys Cox}
\email{srcox@cs.aau.dk}
\orcid{0000-0002-4558-6610}
\affiliation{%
  \institution{Aalborg University}
  \city{Aalborg}
  \country{Denmark}
}

\author{Robert Xiao}
\affiliation{
  \institution{University of British Columbia}
  \city{Vancouver}
  \state{BC}
  \country{Canada} 
}
\email{brx@cs.ubc.ca}
\orcid{0000-0003-4306-8825}


\begin{abstract}
While human--AI collaboration systems have increasingly been built to increase efficiency or support creativity, little work has examined how the design of interactions shapes the social connection between human and artificial agent. We examine how the temporal and visual dimensions of collaboration shape the experience of a writing task. Specifically, we built three variants of an AI-assisted text editor along a spectrum of simulated humanlike interaction (synchronous and with a cursor) to machinelike interaction (asynchronous and without a cursor), and conducted a comparative user study (n=48). Our exploratory findings suggest that synchronous suggestions increased efficiency but led to contextual misalignment, while a visual cursor increased intent understanding but evoked feelings of surveillance. Taken together, humanlike design of artificial agents can create positive social expectations but also elicit social costs, especially without the alignment present in human--human collaboration. We extend our findings into design implications and ethical considerations when building human--AI collaboration systems. 
\end{abstract}




\maketitle

\section{Introduction}

As the capabilities of AI systems rapidly advance, increasing amounts of HCI research have examined the potentials and challenges of human--AI collaboration. Human--AI collaboration, which involves integrating AI into present human working environments \cite{wangHumanHumanCollaborationHumanAI2020, fragiadakisEvaluatingHumanAICollaboration2025}, has been explored and applied across various domains for improving efficiency and productivity \cite{fui-hoonnahGenerativeAIChatGPT2023,cox2025beyond}, spurring creativity \cite{hanWhenTeamsEmbrace2024, huDesigningInteractionsGenerative2025}, and generating feelings of social support \cite{zhengCustomizingEmotionalSupport2025, liLLMUseMental2026}. However, this can come at the expense of human autonomy and emotional reliance \cite{zhengCustomizingEmotionalSupport2025, huDesigningInteractionsGenerative2025}, which loosely tie to basic human needs \cite{deci2012self}. In particular, the social relationship between AI systems and humans is important in shaping such benefits and tensions. Researchers find that users anthropomorphize AI systems \cite{siemonAnthropomorphismSocialPresence2026, schlesenerAmUnderstoodHow2026, basoahNotUsHunty2025a, chaturvediExploringFrontierAnthropomorphism2025} and even form deep relationships \cite{leeNegotiatingRelationshipsChatGPT2026, maPrivacyHumanAIRomantic2026}, as the line between their perception of such systems as being \emph{humanlike} or \emph{machinelike} blurs. The anthropomorphic views that people hold towards AI systems shape the way in which users adopt, trust, and use them \cite{chenPortrayingLargeLanguage2025}. 

Anthropomorphism is one factor that shapes the broader understanding of technology use, an enduring dimension of HCI research. Nass et al. \cite{nassComputersAreSocial1994} proposed the \textbf{Computers are Social Actors} (CASA) theory, highlighting how people have social responses to technology that draw from human--human interaction norms, allowing even primitive humanlike cues to elicit powerful social responses~\cite{nassComputersAreSocial1994, nassAnthropomorphismAgencyEthopoeia1993}. Over time, various cues have been explored for potentially evoking different social responses to technological systems, such as simulated voice~\cite{seabornVoiceHumanAgent2021, nassComputersAreSocial1994}, appearance~\cite{komatsuHowDoesAgents2011a, chenDifferentDimensionsAnthropomorphic2024, liAnthropomorphismBringsUs2021}, personality~\cite{rahmanVibeCheckUnderstanding2026, mathurWhoWantsBe2026}, and sociolects~\cite{basoahNotUsHunty2025a}. Across such studies, researchers show how design cues can shape people's attitudes, trust, and feelings of psychological distance towards technological systems; this has increasingly been relevant in AI contexts with the prevailing incorporation of large language models (LLMs) that respond in humanlike manners.

However, many previously studied cues are \textbf{innate elements} of the system; they represent a consistent underlying quality embedded into the system. Less work has examined the design of \textbf{interaction elements} --- elements that shape individual interactions with technology, such as \textbf{when} interaction occurs and \textbf{how} the interaction is visually presented. While recent research has begun to study the temporal \cite{bucincaTrustThinkCognitive2021a, leeWeWaitHow2025, liuHowAIProcessing2024} and visual quality \cite{cox2026watching, zhouWordsInfusingConversational2024a} of interaction design, our work looks uniquely at how intentionally shaping these interactional dimensions to be more humanlike affects people's experiences in a collaborative task, and their social perception of the AI system (and its underlying \emph{artificial agent}). 

To study this, we built and explored three variants of a human--AI collaborative writing system. Inspired by synchronous peer editors such as Google Docs \cite{GoogleDocs}, our system involves an artificial agent acting as the peer in providing suggestions to user-written content. The variants vary on the spectrum of temporal humanlikeness (whether the changes are presented autonomously during writing or only upon user prompting) and visual humanlikeness (whether the AI needs to move a cursor and type character by character or it makes changes immediately). Using these variants, we conduct a comparative user study (n=48) to explore the singular research question of:

\begin{quote}
    \textbf{RQ}: How does simulated temporal and visual humanlikeness shape experience and social perception in a human--AI collaborative writing task? 
\end{quote}

When the agent made suggestions autonomously and synchronously, participants felt more efficient in embedding suggestions directly into their writing. However, participants also actively monitored the agent for changes and felt distracted, and were furthermore influenced by the agent's suggestions during the ongoing writing process. When the agent had a cursor that moved around while suggesting, participants felt that they had a better understanding of the agent's intention. Yet, this caused them to react to such intention, waiting and anticipating the agent's response. 

Taken together, while humanlike collaborative systems \textbf{do elicit social presence}, this can potentially lead to \textbf{social costs} as well --- participants at times felt monitored by, judged by, and compared against the agent. These costs were exacerbated by insufficient alignment between the human and the artificial agent, compared to interacting with known humans, but could be somewhat alleviated through visible alignment and shared understanding. 
Altogether, we contextualize these findings around prior psychological theory, 
drawing out how social presence in humanlike agents may both support and strain collaboration, and highlighting implications for the design of more visibly aligned collaborative systems.

\section{Related Work}

We consider prior literature to contextualize our research. We examine technology's role beyond its practical outcome, specifying how AI-embedded systems can take on social roles that people relate to and interpret. We consider how design can shape engagement with the artificial agents of such systems, and how research has examined both positive and negative consequences of engagement. Finally, we investigate the role of artificial agents more specifically in a collaborative writing context, and how our work extends the literature on system design in this area.  

\subsection{Technology's Role as Social Agents}

Nass et al.'s \textbf{Computers Are Social Actors} (CASA) theory \cite{nassComputersAreSocial1994} proposes that people interact socially with technology, applying human norms in response to even primitive humanlike cues. 
Researchers have shown that how technological systems interact with people shapes how they are perceived socially; for example, specific choices of human voice incorporated into a computer can strengthen perceptions of it as a social agent \cite{nassAnthropomorphismAgencyEthopoeia1993, leeDesigningSocialPresence2003}.
CASA theory has been widely engaged with, extended, and critiqued over the years. Gambino et al. \cite{gambinoBuildingStrongerCASA2020} highlight that one reason a more nuanced understanding of CASA is required is simply that humans (and the way they interact with technologies) have changed; Lang et al. \cite{langAreComputersStill2013} highlight that desensitization through common use may shift social attitudes towards technological systems. 

With the development of newer technologies such as robots and chatbots, researchers have once again drawn on CASA theory to contextualize digital interactions with such systems. Seok et al. \cite{seokWhatEmotionsPersonalities2025} adopted CASA to explain acceptance of LLMs, highlighting how user personality traits play a role in social perception. Baseman et al. \cite{basemanPokerPlayMoney2025} studied psychotherapists' role-playing with LLM-based virtual patients, demonstrating these technologies' power in eliciting social responses. Cues such as identity and expression can further evoke social responses in interactions with LLM-driven chatbots \cite{liExploringEffectsChatbot2025}. 

One driving factor behind the social responses outlined in CASA is the human tendency to anthropomorphize~\cite{nassAnthropomorphismAgencyEthopoeia1993, xuDeepMindSocial2022}. Anthropomorphization has been heavily discussed in the AI domain~\cite{sallesAnthropomorphismAI2020a, liMachinelikeHumanlikeLiterature2021, liAnthropomorphismBringsUs2021}, especially in regards to LLMs~\cite{siemonAnthropomorphismSocialPresence2026, schlesenerAmUnderstoodHow2026, basoahNotUsHunty2025a, chaturvediExploringFrontierAnthropomorphism2025}. LLMs have become increasingly social and humanlike in their responses, which leads to innate anthropomorphization \cite{iyerWasThereBias2025}; users may personify such systems as friends or romantic partners \cite{liItsHavingFriend2025, maPrivacyHumanAIRomantic2026, leeNegotiatingRelationshipsChatGPT2026}. Such systems can offer companionship, social engagement, and a space for support \cite{liuChatGPTPerspectivesHuman2024}, and are therefore used in both personal~\cite{liItsHavingFriend2025, maPrivacyHumanAIRomantic2026, leeNegotiatingRelationshipsChatGPT2026} and business~\cite{saputraAnthropomorphismbasedArtificialIntelligence2024, israfilzadeRoleGenerativeAI2023, kumarAnthropomorphicGenerativeAI2025, greilichConsumerResponseAnthropomorphism2025} settings.

Yet, anthropomorphism can also lead to a number of well-studied challenges. For instance, research has explored concerns around misplacement of trust and overestimation of humanlike capabilities \cite{jiDemystifyChatGPTAnthropomorphism2024, iyerWasThereBias2025}, as LLM interactions shape mental attributions and subsequent user feelings \cite{colombattoInfluenceMentalState2025, akbulutAllTooHuman2024}. A strong social connection with LLMs can even make hallucinated information seem trustworthy when presented amicably \cite{maedaWhenHumanAIInteractions2024}, and the companionship offered by LLMs can easily sway into emotional attachment and dependency \cite{liDoubleedgedSwordEffect2025, yankouskayaCanChatGPTBe2025}. 

Altogether, humans innately respond to cues and anthropomorphize technological systems, including artificial agents. Our work extends existing CASA by exploring how eliciting humanlikeness through temporal and visual interaction cues during a collaborative writing task shapes the overall experience. We explore how such factors shape social responses and what happens when social expectations are met or mismatched. 

\subsection{The Role of Design for Artificial Agents}

Although humans innately anthropomorphize technology, specific design decisions can exacerbate humanlike perception and social response. For instance, voice \cite{seabornVoiceHumanAgent2021, nassComputersAreSocial1994} and appearance \cite{komatsuHowDoesAgents2011a, chenDifferentDimensionsAnthropomorphic2024, liAnthropomorphismBringsUs2021} are two major design elements that shape responses and feelings towards artificial agents, mediating attitudes towards the agent and the level of psychological distance \cite{liAnthropomorphismBringsUs2021}. Other design elements, such as conversational style and expressivity, also shift attitudes and feelings of social presence during interactions with technology \cite{chenDifferentDimensionsAnthropomorphic2024, basoahNotUsHunty2025a, anejaUnderstandingConversationalExpressive2021a}.  

While personality, appearance, and voice are design qualities inherent to the \emph{agent}, our work explores design qualities inherent to \emph{interactions}. We first consider temporality --- cues that affect \emph{when} interactions are completed. Prior literature on human--chatbot interactions \cite{gnewuchFasterNotAlways2018, shiwaHowQuicklyShould2009} suggests that instantaneous responses by chatbots are less humanlike, affecting user perception of the chatbot as a social agent; this has extended into research on LLMs \cite{tan2026impact, dubielHeyGenieYou2024}. Buçinca et al. \cite{bucincaTrustThinkCognitive2021a} studied how forcing functions, which affect when and how LLM responses are shown, shape more mindful engagement with LLM agents; we draw a comparison between their work and Cox et al.'s design frictions \cite{coxDesignFrictionsMindful2016a}. More mindful interaction (opposed to mindless anthropomorphization) could also potentially affect social presence and CASA response \cite{xuDeepMindSocial2022}. The temporal element of generative AI delay in different contexts can also offer a signal of effort and a period for anticipation \cite{leeWeWaitHow2025}, potentially shaping trust towards the system \cite{liuHowAIProcessing2024}. 

We also consider the visuality of action --- cues that show \emph{how} the interaction is completed. Zhou and Hu \cite{zhouWordsInfusingConversational2024a} highlight how extending the response generation of LLMs with small humanlike typing behaviours (hesitation, self-editing) can make LLM systems feel more natural and trustworthy. Cox et al. \cite{cox2026watching} highlight how visual displays of "thinking" as short reflective statements also shape anthropomorphization of the agent and user expectations. 

Understanding how design shapes experience is important --- design can induce both positive and negative social outcomes. While Li et al. \cite{liExploringEffectsChatbot2025} highlight how human social responses towards generative AI can give rise to prosocial behaviour in certain contexts, Mazhar et al. \cite{mazharGenerativeArtificialIntelligence2026} show that this may also lead to long-term reliance in other contexts. More broadly, we draw a connection between the negative social outcomes of LLMs and the ethically-problematic \textbf{dark patterns}, which use design to undermine a user's best interests \cite{changTheorizingDeceptionScoping2024, grayDarkPatternsSide2018, bongard-blanchyAmDefinitelyManipulated2021}. For instance, Krämer et al. \cite{kramerTrickingTrustingInfluence2025} highlight how the social cues of LLM systems may elicit trust beyond the system's capabilities. This is especially problematic given LLMs' existing biases \cite{rutinowskiSelfPerceptionPoliticalBiases2024a}, and their propensity to shape user attitudes towards social issues when designed in specific ways (e.g. as autocomplete suggestions \cite{williams-ceciBiasedAIWriting2026}). 

LLMs continue to be explored in social roles in domains such as mental health support \cite{nazirChatGPTMentalHealth2026, jungIveTalkedChatGPT2025, girayCasesUsingChatGPT2025}, clinical workflows~\cite{spatscheckEffectsGenerativeAIs2024, liExploringFutureAI2026}, and education \cite{vibergChattingCodeExploring2025, parkPromisePerilChatGPT2024}. Thus, understanding how interaction cues during a task shape social relationships and user attitudes is imperative (which can exist almost independently of the generative content). We use prior knowledge and implications of technological design to contextualize how design cues that elicit visual and temporal humanlikeness shape experiences and attitudes during collaborative writing. 

\subsection{Writing with Other Humans and Artificial Agents}

In 2010, Google introduced a feature to allow people to see real-time edits to Google Docs files from other users~\cite{GoogleDocs}, as a form of human--human collaborative real-time editing \cite{ebadiExploringImpactOnline2017}. In collaborative writing, awareness of others shapes the context of one's own activity \cite{dourish1992awareness}. Liu and Lan highlight how text-based collaboration can support participation and engagement \cite{liuSocialConstructivistApproach2016, ishtaiwaImpactGoogleDocs2015}, creating a social, communal space to share, act, and know \cite{lipponenAssessingCollaborativelyUsable2004, roschelle1992should}. Extending this social basis, Jung et al. find that collaboration on Google Docs shifts the turn-taking nature of social collaboration towards simultaneous working \cite{jungPossibilitiesLimitationsOnline2017}. Such collaboration supported divided participation, shifting social connection and group awareness \cite{jungPossibilitiesLimitationsOnline2017}. Overall, synchronous collaboration tools shape social connectivity and group dynamics in ways that differ from asynchronous interaction \cite{scissorsRealtimeCollaborativeEditing2011, birnholtzTrackingChangesCollaborative2012}.

Rapid improvements in AI capabilities have enabled the development of systems that support writing ``collaboratively'' with AI agents. For instance, VISAR is an AI-enabled writing assistance system that supports the writing process during various steps, helping with brainstorming, organization, and argumentation \cite{zhangVISARHumanAIArgumentative2023a}. Lehmann et al. developed a collaborative text editor that involves multiple users and artificial agents \cite{lehmannCollaborativeDocumentEditing2026a}. They considered how writers create and interact with these agents, finding that authors both supervise and delegate to their agents \cite{lehmannCollaborativeDocumentEditing2026a}. At a broader level, Reza et al. conducted a literature review on AI interventions in writing processes \cite{rezaCoWritingAIHuman2025a}, highlighting prior strategies for writing support and their effects on agency and ownership.

Since the public release of LLMs, turn-based prompting has been the dominant form of interaction \cite{gaoTaxonomyHumanLLMInteraction2024a}. In accordance, AI writing-support tools often mimic this turn-based creation paradigm \cite{rezaCoWritingAIHuman2025a, mysorePrototypicalHumanAICollaboration2025, kimDiaryMateUnderstandingUser2024b} instead of providing the real-time collaborative editing interactions. While some research has explored synchronous collaboration, they primarily focus on system function and usability rather than design \cite{lehmannCollaborativeDocumentEditing2026a}. Yet, design is important --- even subtle factors such as typing speed \cite{zhouThoughtfulConfusedUntrustworthy2025a}, intrusiveness of suggestions \cite{yinProactiveAICatalyst2026}, and the hierarchical level of comments \cite{dhillonShapingHumanAICollaboration2024a}, all shape the experience of AI-assisted writing. As such, there is an evident gap in design and dynamics between human--human collaboration and human--AI collaboration in writing, motivating our choice of the domain. Research can more deeply consider the impact of design during collaboration with an artificial agent (and their potential social consequences); we explore this gap in our work.  


\section{System Implementation}

Our research question considers the effects of two independent variables of humanlike agentic behaviour --- \textbf{temporal humanlikeness} and \textbf{visual humanlikeness} --- in collaborative tasks. In particular, our research specifically applies these variables to the specific task of peer editing, in which humans are tasked with creating a written response to a prompt, and an artificial agent aims to help them by providing subjective suggestions for improvement. This was inspired by collaborative text editing processes on existing platforms such as Google Docs or Overleaf \cite{ebadiExploringImpactOnline2017}. 

Thus, we built a custom text editor that was hosted online. In this section, we discuss how we augmented this base text editor with an artificial agent acting as a peer editor. This agent made suggestions for the user's written text in three different ways (\emph{three variants}) that span the spectrum of humanlikeness in interaction. 

\subsection{Three Variants of an AI-Assisted Text Editor}

\subsubsection{Types of Suggestions}

\begin{figure*}[h]
\centering
\includegraphics[width=1.0\textwidth]{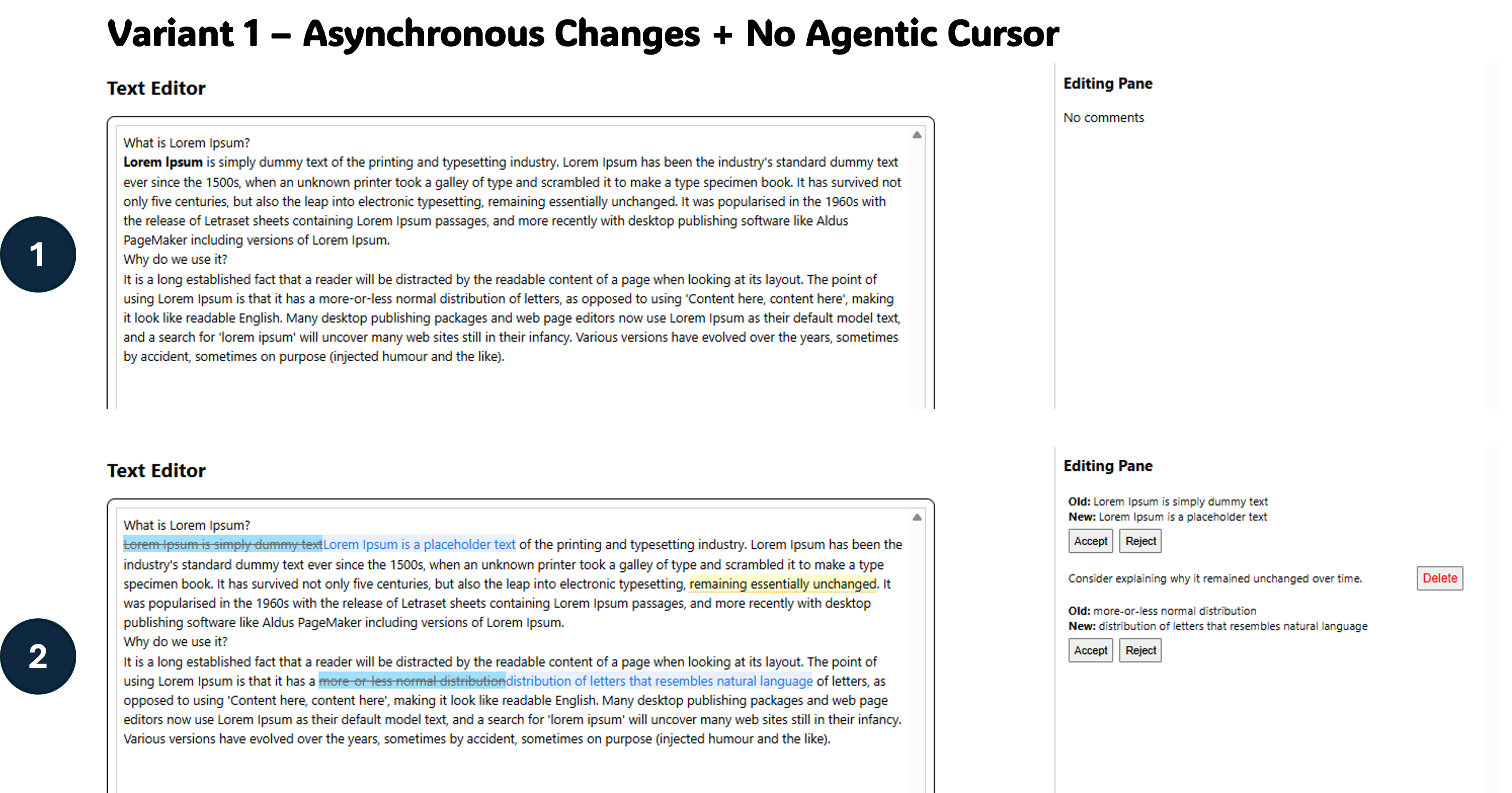}
\caption{Variant 1 of our text editor. In (1), the author writes out the text that they would like feedback on, and upon pressing a button (not visible, but under the text box), the AI provides feedback, i.e., seen in (2).}
\Description{This image shows the flow of variant 1 of our system. In subimage (1), we see that the author has written a few paragraphs in the text box and that at present, the editing pane is empty. Upon pressing a button to receive AI suggestions, in subimage (2) we see that the relevant suggestions in the text are highlighted by colour, and are associated with comments in the editing pane (either suggestions for replacements or comments on the text)}
\label{fig:variant1}
\end{figure*}

\begin{figure*}[h]
\centering
\includegraphics[width=1.0\textwidth]{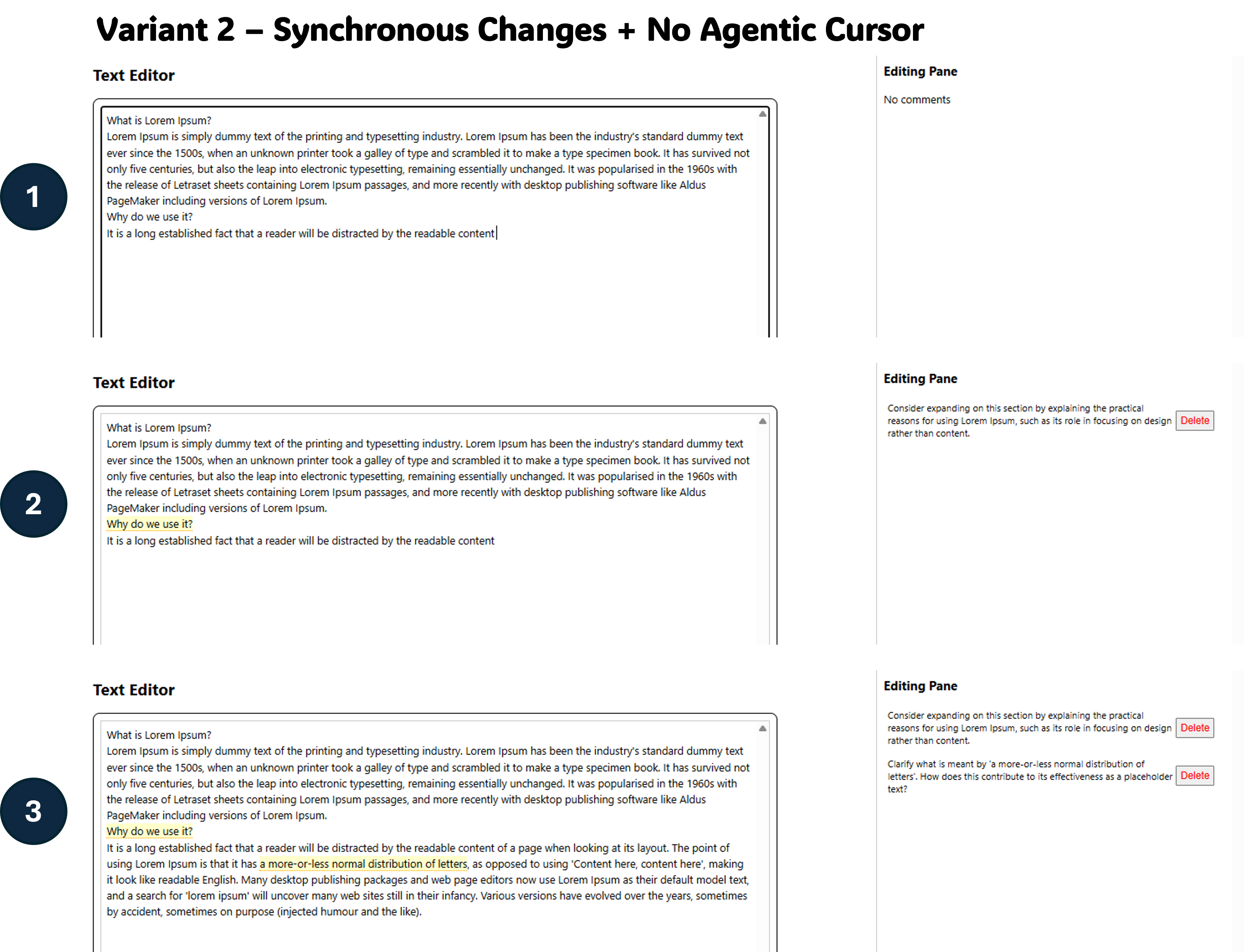}
\caption{Variant 2 of our text editor. While the author is writing, as in (1), the AI makes suggestions along the way. These suggestions (i.e., the highlighted text and comment box) appear instantaneously, such as the comment suggestions in (2) and (3).}
\Description{This image shows the flow of variant 2 of our system. In subimage (1), we see that the author is writing text in the text box, and the editing pane is presently empty. As the author writes, the AI synchronously makes a comment suggestion in (2), noting that the suggestion is made completely and instantaneously. The author continues to write, and the AI makes another comment suggestion in (3), again made completely instantaneously.}
\label{fig:variant2}
\end{figure*}

\begin{figure*}[h]
\centering
\includegraphics[width=1.0\textwidth]{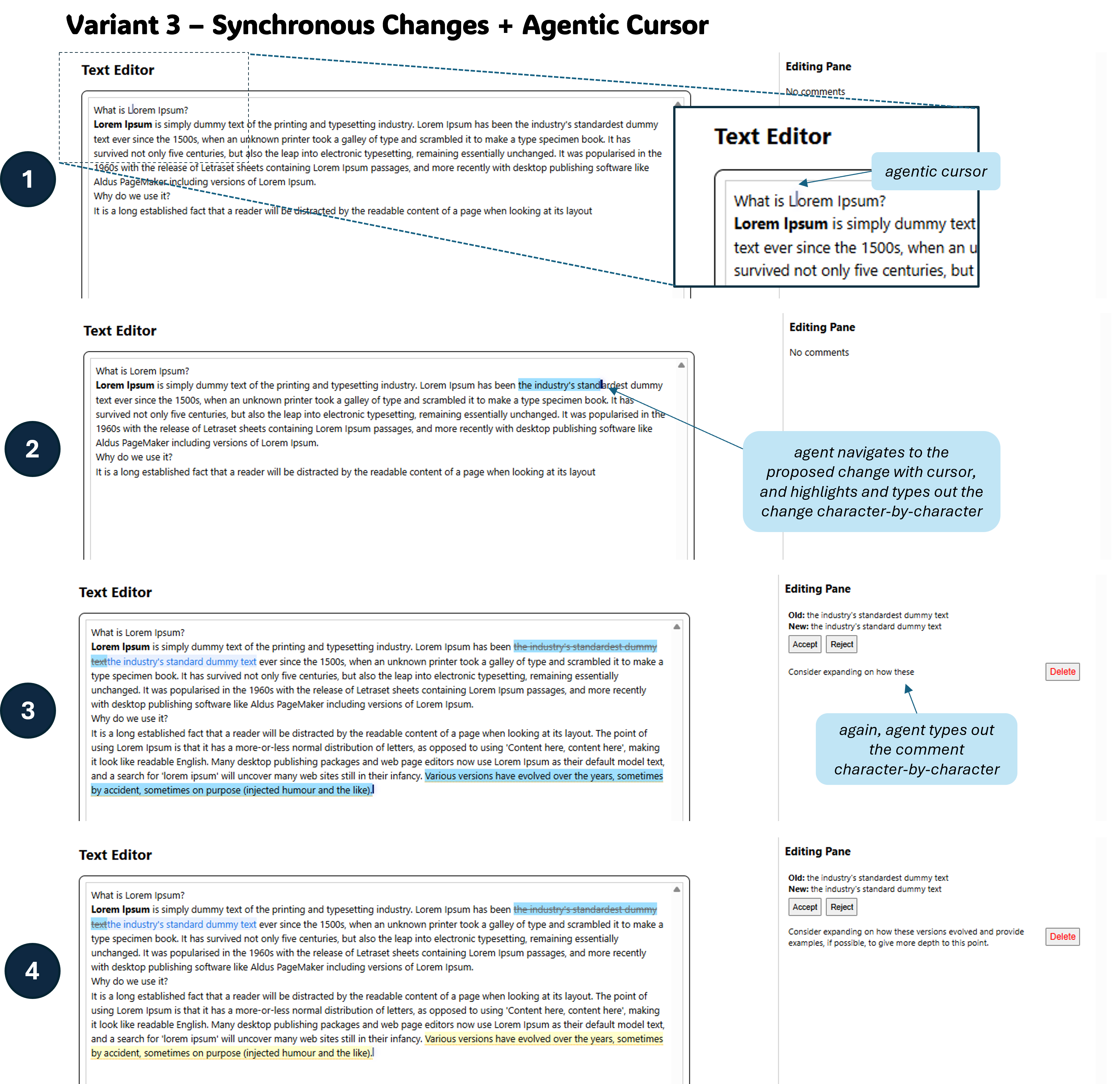}
\caption{Variant 3 of our text editor. While the author is writing, as in (1), the AI has an agentic cursor that it moves within the text. While the author writes, the AI decides to suggest (2). The AI navigates its cursor to the proposed suggestion location and both highlights and types out the suggestion character by character, much like a human would need to. In (3), we see again how the AI needs to type out a comment character by character as well, with all the suggestions seen in totality in (4).}
\Description{This image shows the flow of variant 3 of our system. In subimage (1), we see that the author is writing text in the text box, and the editing pane is presently empty. There is a zoom-in on the text, where we see that an agentic cursor, i.e. a flashing cursor, exists in the text as well. As the author writes, the cursor moves. In (2), the AI synchronously decides to make a replacement suggestion, moving its cursor to the correct location of the suggestion, highlighting the replaced text character by character, and typing out the replacement text character by character. In (3), the AI synchronously decides to make a comment suggestion, once again moving its cursor to the correct location and typing out the comment character by character. In (4), we see the final text editor, involving a text box with highlighted changes in-text and an editing pane with suggested comments. }
\label{fig:variant3}
\end{figure*}

While the artificial agent behaves differently in \emph{when} and \emph{how} it makes suggestions, the actual suite of suggestions in all variants was always constant. In particular, there are two types of suggestions, which are represented in an \textbf{Editing Pane} on the right of the textbox (the \textbf{Text Editor}) --- see Figures~\ref{fig:variant1}, \ref{fig:variant2}, and \ref{fig:variant3} for examples of our overall text editor and the two types of suggestions: 

\begin{itemize}
    \item \textbf{Replacement}: A suggestion to replace a phrase or sentence with an improved version. The text to be replaced is highlighted and struck out, while the new suggested text is inserted after in a new text and highlight colour. This is akin to how \emph{Suggestion Mode} works on Google Docs; similarly, the human writer can choose to \emph{accept} or \emph{reject} the suggestion by clicking a button in the Editing Pane. 
    \item \textbf{Comment}: A suggestion to attach a comment to a part of the text. The comment appears in a textbox in the Editing Pane. This is akin to how \emph{Comment Mode} works on Google Docs. The human writer can choose to \emph{delete} the comment if they wish to, by clicking on a button in the Editing Pane.  
\end{itemize}

\subsubsection{Mapping Independent Variables}

Next, we highlight how we integrate our independent variables into the design of a standard text editor, which we use in a collaborative writing task. 

\begin{itemize}
    \item \textbf{\underline{\smash{Temporal Humanlikeness}}: Autonomous Action and Collaboration Synchrony}: 
    In our study, we position temporal humanlikeness as a binary variable, contrasting synchronous, autonomous collaboration that occurs while the user is writing with asynchronous, user-initiated collaboration that occurs only under the user's control.
    \begin{itemize}
        \item In conditions involving \emph{temporally humanlike collaboration}, the agent works alongside the human in real time, autonomously suggesting text as the user writes without needing to be prompted.
        \item In conditions involving \emph{temporally nonhumanlike collaboration}, the agent works separately from the human writing process, only suggesting text when the user stops writing and explicitly prompts it for suggestions and feedback.
    \end{itemize}

    \item \textbf{\underline{\smash{Visual Humanlikeness}}: Apparent Effort and an Agentic Cursor}: 
    We operationalize visual humanlikeness through an agentic cursor, which visually represents the agent's apparent effort during collaboration.
    \begin{itemize}
        \item In conditions involving \emph{visually humanlike collaboration}, the interface includes an \textbf{agentic cursor} that makes the agent appear to navigate the text using a cursor as a human would. For instance, to make a suggestion, the agent appears to navigate its cursor to the correct location, highlight the relevant text character by character, and type the replacement text character by character. In this way, it mimics the visible inefficiency and effort of human editing, constrained by typing and cursor navigation. Furthermore, this condition also conveys the humanlike effort of reading, as the cursor idly moves forward, fidgets, and teleports even when not suggesting (discussed further in §~\ref{sec:heuristics}). 
        \item In conditions involving \emph{visually nonhumanlike collaboration}, the interface does not include an agentic cursor. Instead, the agent's suggestions appear instantly.
    \end{itemize}
    
\end{itemize}

Given the construction of these design features that serve as representations of our independent variables, we consequently develop 3 variants of the text editor: 



\begin{center}
\begin{tabular}{@{}l l c l l@{}}
$\bullet$ \textbf{\underline{\smash{Variant 1}}} & Temporally Nonhumanlike & / & Visually Nonhumanlike & : Figure \ref{fig:variant1} \\
$\bullet$ \textbf{\underline{\smash{Variant 2}}} & Temporally Humanlike    & / & Visually Nonhumanlike & : Figure \ref{fig:variant2} \\
$\bullet$ \textbf{\underline{\smash{Variant 3}}} & Temporally Humanlike    & / & Visually Humanlike    & : Figure \ref{fig:variant3}
\end{tabular}
\end{center}

We did not include a condition combining temporal nonhumanlikeness with visual humanlikeness, as this combination was conceptually inconsistent with the logic of our manipulation
--- the agentic cursor is intended to visualize humanlike effort during the peer editing process as it unfolds \emph{over time}, whereas asynchronous, user-prompted suggestions occur at a discrete moment. Overall, these variants represent a spectrum of the most machinelike behaviour (\emph{Variant 1}) to the most humanlike behaviour (\emph{Variant 3}). 


\subsection{Developing Heuristics Through Formative Studies}
\label{sec:heuristics}

The task of mimicking humanlikeness --- both temporally regarding the frequency of autonomous action and visually regarding the actions for apparent effort --- using available digital tools required us to develop \emph{heuristics} of humanlike behaviour, e.g., how often a suggestion should be made, how the cursor should move, and so forth. To imbue a machine with humanlike behaviour, we first needed to understand how humans themselves operate during collaboration. 

We performed a short, small-scale formative study (approved by our ethics board) with 8 colleagues (all HCI researchers). In this formative study, the researcher would write out a standard prepared essay (on the prompt: ``\textit{Write a short opinion piece to describe your views on how smartphones affect people's daily lives}''\footnote{This is a modernized interpretation of the first prompt found in an ASAP dataset \cite{mathiasASAPEnrichingASAP2018a}}), and the participants would serve the role as the collaborative peer editor, the eventual role given to the artificial agent in our system. We collected participants' interactions on the page and performed a brief interview about their decisions during collaboration. 

This formative study informed several design decisions. For instance, our original consideration of the system included having the agent make changes directly on the text rather than suggestions. However, participants generally did not change text directly, citing politeness and not wanting to encroach on the writer's content. By timing participant interactions, we were also able to understand the general distribution of when changes were made and the average time between suggestions. We were also able to broadly understand participants' general cursor movement, as most participants moved their cursor sentence-by-sentence as they read, while also jumping back and forth or fidgeting. Beyond this formative study, we also drew upon heuristics within the literature (e.g. the speed and variance at which the agent ``typed'' was informed by the interkey interval of a fast typer \cite{dhakalObservationsTyping1362018}) as well as our own judgment of what heuristics were appropriate (e.g. clamping the rate of suggestions to never be too fast or too slow, etc.)

Although we observed these behaviours and tried to encode them into our humanlike AI through guided rules to the best of our abilities (e.g. would randomly select what type of motion to do based on a defined distribution, at what frequency, etc.), we ultimately acknowledge limitations. Firstly, these were tweaked subjectively until the researcher felt they seemed humanlike when compared to the observations in the formative studies, and secondly, taking an average approximation (e.g. the average time between suggestions) is not always humanlike --- there is no "average" human. 

\subsection{Technical Implementation}

Our text editor was developed and hosted on the Internet, accessible through a web browser. The front-end was created using the React library, in JavaScript, HTML, and CSS. In particular, we used the open-source ProseMirror toolkit\footnote{\url{https://prosemirror.net/}} for creating and adding custom functionality to the text editor. Our system's back-end, which managed data logging and was responsible for OpenAI (gpt-4o) API calls, was developed in Python using the Flask\footnote{\url{https://flask.palletsprojects.com/en/stable/}} library. These API calls took an input JSON object of (1) the present written text, chunked by paragraph, and (2) the previous suggestions, to return new suggestions for the text. Specific prompts used for these API calls can be found in Appendix \ref{appendix:prompts}. 

\section{User Study}

\subsection{Study Protocol}

In designing the user study, we chose to perform a primarily between-subjects design over a within-subjects design for two main reasons: (1) practicality, due to the excessive effort and time that it would take for participants to complete a writing task 3 times, and (2) potentially associated carryover effects, coming from needing to maintain a similar degree of initial interest and engagement across 3 conditions. However, one potential limitation of a purely between-subjects design is that it lacks the richness of comparison that comes from experiencing several conditions. As a result, we adjusted our study accordingly, and each user study session involved: 

\begin{itemize}
    \item \textbf{A main writing task.} Participants were provided a standard writing prompt used in the formative study (i.e., ``\textit{Write a short opinion piece to describe your views on how smartphones affect people's daily lives}'') and instructed to write an essay based on this prompt, with the agent collaborating on the task. Although up to their own interpretation and writing style, we mentioned that participants could simply write a standard essay split into multiple paragraphs. The writing task continued until the participants finished, but was capped at approximately 25 minutes. Participants' experience of the main writing task was captured and compared through both quantitative and qualitative data. 
    \item \textbf{Two subtasks.} After the main writing task, users were given two smaller writing prompts to explore with the remaining two variants (i.e. subtask 1 was to write about ``\textit{any positive encounter / event you had recently}'', and subtask 2 was to write about ``\textit{any attended event, recently or otherwise}''). We asked participants to write approximately one paragraph length. The focus of these subtasks was not on the content, but on exploring how different presentations of humanlikeness might potentially shape the writing experience. Thus, we provided low-stakes, personal prompts that would be easy to write quickly. Participants were asked to explore and speculate on comparison, and this data was captured purely as qualitative data. Exploration of these subtasks was more flexible based on participant responses, but generally took around 10 minutes each. 
\end{itemize}

Based on this study design, we captured between-subject quantitative data from standardized scales performed after the main writing task. We used the \textbf{Partner Modelling Questionnaire} (\textbf{PMQ}) \cite{doylePartnerModellingQuestionnaire2025a} and the short version of the \textbf{Artificial-Social-Agent Questionnaire} (\textbf{ASA}) \cite{fitrianieArtificialsocialagentQuestionnaireEstablishing2022a}, these scales query the participant's perception of the artificial agent, especially their capabilities as partners and social agents, tying back to the theoretical lens of CASA \cite{nassComputersAreSocial1994}. Our study also captures within-subject qualitative data from short interviews after trying each variant. We asked open-ended questions about what each participant liked and disliked about each variant, how it compared to the prior variants they had seen, how the humanlikeness shaped their experience, and so forth. The interview script can be found in the supplemental material. 

Across participants, the main variant was pseudo-randomly assigned (to maintain an equal number of participants across variants); the remaining variants were counterbalanced for the subtasks. Our study received approval from our institute's ethical review board. Studies took approximately one hour, and participants were compensated \$16 CAD.

\subsection{Participants}
We recruited a sample of 48 total participants through postings made on external Slack channels or our institute's paid studies listing page; this resulted in 16 participants for each variant for the main writing condition. The eligibility criteria were to be 18 years of age or older and to be fluent in English. We completed recruitment when we subjectively gauged we had reached a sufficient level of information power \cite{malterudSampleSizeQualitative2016a} in the qualitative data, which was evaluated on an ongoing basis in discussions between the researchers.  

Based on our demographic questionnaire, our study consisted of 34 women, 13 men, and 1 non-binary participant, with an average age of 28.1 (ranging between 19 and 54). Participants generally had writing experience in several contexts --- the most frequent were academic (e.g. papers, essays; n=31) or professional (e.g. emails, proposals; n=39) contexts, but participants also wrote for creative reasons, for documentation, and as personal notes. 

Regarding the frequency of using collaborative text editors to work with others, 1 participant reported essentially never using them, 9 reported rare use (a few times a year), 14 reported occasional use (about once a month), 10 reported frequent use (a few times a month), and 14 reported very frequent use (weekly or more). Regarding using generative AI for writing support, 6 reported never using generative AI, 5 reported rare use, 13 reported occasional use, 9 reported frequent use, and 15 reported very frequent use (along the same frequency labels). Before the study, participants read, reviewed, and signed a consent form regarding data collection and usage. We also provided the prompt of the main writing task to mitigate against ``blanking out'' during the study, but reiterated that no preparatory work was needed.

\subsection{Data Analysis}

Due to the fact that our sample was determined by our qualitative data, which provides richer exploratory insights, we primarily lean on our interview data in our findings, yet use our quantitative data to triangulate trends. Our qualitative data was analyzed through a thematic analysis approach \cite{clarkeThematicAnalysis2017a, braun2021TA}. Based on Braun and Clarke's formulation of thematic analysis, we took a flexible approach in coding the data, interpreting meaning from participant responses and iterating on our analysis throughout the process. Codes constantly shifted, merged, and evolved during the coding process. After coding, we mapped the codes into broader categories through affinity diagramming and visual mapping (see supplemental material). This process helped the researcher develop themes, grounded in the relationships and groupings of the codes. While the coding and analysis were primarily performed by a single researcher, ongoing discussion with the research team helped provide alternative perspectives and guide the formation of the themes. 

In analyzing the quantitative data, we obtained the ASA-scores \cite{fitrianieArtificialsocialagentQuestionnaireEstablishing2022a} and the PMQ scores \cite{doylePartnerModellingQuestionnaire2025a} (the total score, plus each of the individual subscale scores). We performed exploratory data analysis, charting out the data in graphical form and investigating descriptive statistics. We then conducted non-parametric Mann-Whitney tests. While we report the values from all three pairwise comparisons, we focus on exploring the effects of temporal humanlikeness (between Variant 1 and Variant 2) and visual humanlikeness (between Variant 2 and Variant 3). Non-parametric tests were chosen due to the relatively small sample size, and the raw data is in the supplemental material. 


\section{Qualitative Findings}
In structuring our findings, we first consider the specific effects of each of the two dimensions of humanlikeness: temporal and visual. We then take a broader perspective, considering how both of them together shape social perceptions during the collaborative writing task, which affects the overall experience of writing. 

\subsection{The Effect of Synchrony --- Temporal Humanlikeness}

We first begin with considering the effects of temporal humanlikeness --- what happened when the AI system autonomously made suggestions on an ongoing basis (synchronously) as the participant was performing the writing task? Participants generally agreed that this felt more human, e.g. \emph{``it felt more human for sure''} (P33), \emph{``the synchronous is almost humanlike''} (P5). Participants also viewed such suggestions as being more collaborative, e.g. \emph{``that's sort of what people could call true collaboration, like your heads are in it together and you're taking the information as you go''} (P23), compared to the more tool-like asynchronous suggestions. 

\subsubsection{Context and Intention --- Missing the Full Picture}

One challenge in synchronous collaboration with an artificial agent was that participants often found that these agents missed the full context of what they wanted to write. Writing is not a linear process in which the participant solely types out words --- they need to consider, plan, and think about \emph{what} they want to write. Yet, because the agent solely understands the words written in the text box, it can lack alignment in terms of the entire process, in particular, missing the context of the participant's planning phase. For instance, P17 and P28 both pointed out this lack of alignment in a synchronous context:

\begin{quote}
    \emph{``That's where I would get confused... I wouldn't think that suggestion was necessary because I was gonna write it anyway''} --- P28
\end{quote}

\begin{quote}
    \emph{``[Variant 2] lacks context, but [Variant 1], I think maybe after reading the entire paragraph, then the AI has a sense of what I'm trying to communicate''} --- P17
\end{quote}

P17 described how the asynchronous system is better aligned with context and intent because it only makes suggestions during user-decided logical breaks in the process (e.g. after each paragraph). Moreover, participants highlighted how the writing process was often non-linear, iterative, and split into separate tasks (e.g. of drafting and then editing). When the agent made suggestions automatically without a full understanding of each participant's process, participants felt like this seemed like \emph{``it was almost like putting words into my mouth or putting suggestions of the future into this''} (P2) or \emph{``it eagerly makes those suggestions''} (P12). Synchronous collaboration amplifies contextual misalignment when artificial agents solely look at ongoing output without understanding the full thinking process. 

\subsubsection{Immediate Feedback and Mental Overload}
Participants found that immediate feedback offered synchronously and autonomously by the artificial agent can have benefits in efficiency. By considering and integrating AI suggestions as an ongoing process during writing, it supports both editing and formulating plans without iteration, merging the aforementioned separate tasks of writing into a continuous process: 

\begin{quote}
    \emph{``With the changes being simultaneously made as you're typing... You don't actually have to always read over the essay''} --- P3
\end{quote}

\begin{quote}
    \emph{``The comments are coming up as you're thinking about things... it makes it easier to embed the suggestions because you're already thinking about whatever you're writing''} --- P4
\end{quote}

Thus, as participants focussed on constructing their subsequent words, they also considered the edits on prior text and suggestions for future changes made by the artificial agent. Some participants viewed this as making the writing process more efficient, e.g. \emph{``it's more efficient... get all of the editing out of the way as you're writing''} (P37), \emph{``[suggestions] being done at the same time... can make things a little bit quicker and a little bit more efficient''} (P45). This was particularly relevant for the replacement suggestions, which were (1) less cognitively demanding to implement, and (2) more aligned with their trust in AI capabilities. Beyond efficiency, P25 commented on how \emph{``humans naturally like instantaneous feedback''}, in that it provides a sense of \emph{``security''}. 

Yet merging the separate processes of conceptualizing, writing, and editing into a singular pass also increases the mental overload and disrupts people from their ongoing task. Obtaining immediate, synchronous feedback led to many participants indicating that it was distracting compared to asynchronous feedback. 

\begin{quote}
    \emph{``If [AI suggestions] pops in, it's kind of distracting to see what it is saying''} --- P22
\end{quote}

\begin{quote}
    \emph{``I had to look at it, and then sometimes I lose my chain of thoughts in the writing. If I write a full paragraph, the whole thing first, and then it gives me suggestions afterwards, it might be better''} --- P20
\end{quote}

Participants remarked on how the visual pop-up associated with the synchronous suggestions took them out of the flow of writing, and that it was difficult to ignore. Some participants indicated that they wanted to quickly remove these visual distractions, e.g. \emph{``I don't want to be disrupted by these highlights during my writing''} (P1), making changes rapidly so that they can dismiss these disruptions. Overall, while synchronous collaboration helped participants perform several subtasks involved in writing on a more efficient and ongoing basis, this also led to visual and mental overload that could take them out of their desired writing flow. 

\subsubsection{Influences and Shapes the Subsequent Writing}

During the synchronous collaboration conditions, the artificial agent offered suggestions on an ongoing basis before the writing was complete. Participants mentioned that such suggestions (in particular, the higher level \textbf{comments}) could shape subsequent writing. This was sometimes viewed positively, as a way of garnering potential perspectives, alternatives, and improvements on what to write:

\begin{quote}
    \emph{``I could like improve it as I'm going along... it's kind of like opening a dialogue on how I could improve my writing in the moment''} --- P38
\end{quote}

\begin{quote}
    \emph{``That was helpful in generating new avenues that I could go through''} --- P5
\end{quote}

In particular, some participants mentioned that getting new ideas to help with writing would be especially helpful if they felt stuck; autonomous, synchronous suggestions could easily offer direction on what to write. P27 mentions that reading through such suggestions while they were writing \emph{``helped me form an idea of what else to say''}. 

However, participants also worried that this influence on subsequent writing could take away their personal voice and their ideas. Participants hypothesized that AI suggestions would affect their writing style and arguments, even if they tried to avoid this. 

\begin{quote}
    \emph{``if I incorporate the first comments, then I feel like my train of thought would change according to its response... it's sort of a bit restrictive in the sense where what you are writing could be guided by AI instead of using AI as a tool to help refine''} --- P14
\end{quote}

\begin{quote}
    \emph{``it could potentially have too much influence on my argument, which is maybe also a bad thing because ultimately the writing activity is for humans to communicate our ideas''} --- P1
\end{quote}

As a result, synchronous suggestions shape the ongoing process of writing. Participants mentioned that this influence could be beneficial when they themselves lacked ideas to shape their immediate arguments, but disliked being influenced when they had strong personal ideas that they wanted to put into writing. For instance, P2 highlights how a combination of synchronous and asynchronous suggestions could be useful depending on whether one needs initial assistance \emph{``to brainstorm and get more ideas''} or when one is writing down their own ideas and \emph{``just want it to stay quiet''}.

\subsection{The Effect of an Agentic Cursor --- Visual Humanlikeness}

We consider the effects of visual humanlikeness --- what happened when the AI system visually had to represent apparent effort through navigation and motion of a cursor to simulate reading and editing? Participants once again agreed that visual humanlikeness through the cursor felt more human and more of a collaborative partner. For instance, P42 described how \emph{``it definitely felt more like a collaborator with [Variant 3]''}, and several participants drew comparisons with collaborative synchronous editing using existing systems, e.g. \emph{``it felt like I was in Google Docs writing with someone else''} (P48) and \emph{``I've used Google Docs and Microsoft Word before, and that's what it looks like when another person is editing the document... I know it's AI, but without that, I think that it's a real person''} (P20). 

However, some participants were minimally affected by the agentic cursor. P24 mentioned that \emph{``that doesn't affect me, the cursor of the AI doesn't affect my cursor''}, and P41 stated that \emph{``in the beginning, I noticed it, and then after that initial jarring [moment]... my brain completely started ignoring it''}. We hypothesize that this may be because while temporal humanlikeness fundamentally changes the process of writing and how subtasks (planning, writing, editing) are organized, visual humanlikeness by itself is a factor that primarily affects perception over the writing process. 

\subsubsection{Distractions and Expectations During Writing}

Similar to the finding in temporal humanlikeness, the artificial agent's cursor also contributed to feelings of distraction during writing. Participants discussed how seeing a cursor moving around visually and making suggestions as they wrote was generally disruptive, noisy, and could cause them to lose their train of thought. 

\begin{quote}
    \emph{``suddenly there's things flicking and things on the screen... it's more of the visual distraction''} --- P25
\end{quote}

\begin{quote}
    \emph{``the moving around while you write is very destabilizing because you're looking at it, and then I lost my train of thought''} --- P40
\end{quote}

In particular, this distraction could come from the sense of anticipation and social presence \cite{kreijnsSocialPresenceConceptualization2022} that the cursor entailed. The existence of the agentic cursor gave participants the impression that the AI was reading and was going to make a suggestion; P4 mentioned that \emph{``it's almost like it's thinking''} and that \emph{``I kind of almost want to wait for it to say something''}, stopping their writing in place while waiting for the agent. This created a social give-and-take, where participants were almost having an unspoken dialogue grounded in anticipation of what the agent seemed like it was going to do:

\begin{quote}
    \emph{``it was distracting, I felt like I was like, `oh, what are they going to suggest now?'''} --- P6
\end{quote}

\begin{quote}
    \emph{``I got a bit nervous to finish the sentence because I didn't really know what was going to happen next, so I kind of paused at an unfinished sentence to see what was happening''} --- P48
\end{quote}

As a result, the existence of an agentic cursor created an almost nonverbal dialogical interaction between the human and the artificial agent drawn from anticipation and expectations of agent behaviour. However, these expectations could end up being distracting, annoying, and elicit hesitation without proper alignment. 

\subsubsection{Understandability of Action --- Aligning to the AI}

However, one benefit of having an agentic cursor during collaborative writing was in providing increased understandability and predictability of the agent's actions. For instance, P15 and P28 both highlighted how it is \emph{``useful''} to know \emph{``what exactly [the AI] is working on''} (P15) and \emph{``where the AI is checking''} (P28). These dimensions were viewed positively, as they could help shape the participant's actions during the writing process:

\begin{quote}
    \emph{``I can know it's making changes in that section and then I can continue writing in this section''} --- P46
\end{quote}

Furthermore, this understandability also helped formulate a sense of connection more akin to collaborating with a human:

\begin{quote}
    \emph{``you know where it's at... it's like there's a human typing. So that is way different from anything I've interacted with, it kind of gives you that connection''} --- P31
\end{quote}

In particular, we found that increased understandability of the AI through visual humanlikeness contributed towards its perception of being more intelligent and collaborative. By understanding how the artificial agent is ``thinking'', we found that this supports the human in actively aligning themselves towards the agent's action:

\begin{quote}
    \emph{``having the cursor there makes it easier to know where the AI is reviewing stuff as opposed to not having it. And then you kind of have to guess how long you have to wait or where it's reading the stuff at''} --- P46
\end{quote}

\begin{quote}
    \emph{``it's reading through the entire text, even though I believe in the previous version [Variant 2], it's also definitely reading through the entire text, right?''} --- P1
\end{quote}

We note how the agentic cursor can be helpful to participants --- the \textbf{human} writers --- because it provides them with the desired awareness and understanding of an artificial agent they otherwise cannot communicate with. As a result, this supports them in adapting their thinking and process, \emph{even if} the understandability of the agent is purely simulated through a visual artifact (highlighted by P1's prior quote). 

\subsection{Collaborating on a Spectrum of Humanlikeness}

In the prior findings, participants felt a sense of social presence from the artificial agent arising from both temporal and visual humanlikeness. Participants, at times, found the system annoying, collaborative, eager, nerve-inducing, and eerie. The writing became almost dialogic under humanlikeness: the agent's suggestions shaped ongoing human writing, and the human sometimes waited for the agent to make these suggestions. In this section, we broaden our discussion by highlighting how this social presence shaped users' perceptions and their relationships with the artificial agent. 

\subsubsection{A Social Presence, for Better or Worse}

One common thread was that having a social presence from humanlike design led to a constant feeling of being watched and judged for writing, e.g. \emph{``I feel like somebody was watching me''} (P2), \emph{``it's like watching me''} (P29), \emph{``I feel like it's monitoring me''} (P32). This constant feeling of being imminently watched was viewed negatively:

\begin{quote}
    \emph{``I don't know if the word is creepy, but yeah, it was like, oh my god, something's going on, someone is typing at the same time that I'm typing. So, I felt, yeah, it made me feel a little bit uncomfortable''} --- P22
\end{quote}

\begin{quote}
    \emph{``So it felt a bit eerie, like someone else is reading my text.''} --- P35
\end{quote}

This could impact the experience of writing as participants feared judgment --- P27 mentioned how it would be \emph{``upsetting''} to get constant suggestions. P34 felt more hesitant to put their unfiltered thoughts on the work (breaking their usual flow) because the AI would make premature comments autonomously, and P1 felt more rushed than usual because it felt like the agentic cursor was always chasing their writing: 

\begin{quote}
    \emph{``I just feel like I have to like really, really think hard about what I'm like writing down instead of kind of like casually just like getting my ideas out... that kind of stops me from just going with the flow''} --- P34
\end{quote}

\begin{quote}
    \emph{``In [Variant 3]... it does feel like something / someone is chasing behind my cursor. I think I need to type faster''} --- P1
\end{quote}

P21 highlighted that with a more humanlike artificial agent, they feel like they form a \emph{``relationship''} with the agent. Oftentimes, we found that this social relationship came with initial tensions that involved participants feeling judged and watched by the agent, and were therefore loath to write in their regular process. As P7 mentioned, \emph{``I compare myself to [the agent] and be like... you're so much faster at typing than me, you can move the cursor so fast''}. 

However, some participants described how a more humanlike agent could elicit the positive \emph{``social aspect''} (P31) of writing that comes with social connectivity. For instance, through appearing more humanlike and seemingly putting in more effort into reading and making suggestions, P41 was much more receptive to the AI suggestions, where if \emph{``I felt like no effort was put into the edits that I was receiving, I feel a bit like angry''} (P41).  

\subsubsection{Alignment with Humans / Misalignment with AI}

This tension from social presence was not unique to collaboration with artificial agents, as some participants drew comparisons to even collaboration with other humans: \emph{``it was just like having a stranger do this with me and I didn't really like that feeling''} (P13). Even with known human authors, this alignment would not always be perfect, and participants may prefer asynchronous collaboration, or at least synchronous collaboration, but away from ongoing writing sections:

\begin{quote}
    \emph{``When I'm doing group projects, I generally prefer to put other things down and then just like... if you want to change them then go ahead and do that, rather than having someone hover over me while I'm typing in the moment''} --- P34
\end{quote}

However, participants still brought up fundamental differences between the experience and feelings of collaboration between a human and an AI, based primarily on trust and connection. For instance, P35 brought up that the human connection comes from making jokes inline during collaboration:

\begin{quote}
    \emph{``being able to make those jokes... I think that's what I was kind of missing, that kind of element of human interaction and human connection, because an AI can approximate that connection, but it kind of missed it for me here. I think that's what makes collaboration fun and engaging, and we would feel more inclined to follow other people's ideas''} --- P35
\end{quote}

This human connection could also come from channels of communication, e.g. \emph{``usually we're on a call and there's a discussion about what's about to be written''} (P3), or just a \emph{``sixth sense... you get a feel for people's cadence and typing''} (P27). When lacking an established human connection, even when the AI is designed to simulate humanlike behaviour, P35 mentioned that \emph{``it was easier for ignoring [the agent] because there was nobody else's emotion to consider''}. 

Participants also drew from a priori knowledge of present AI systems (especially LLMs) and their capabilities when evaluating their collaborative role. P3 discussed how the humanlike system \emph{``mimics human behaviour, but doesn't actually know what I'm necessarily thinking or... my desires''}; highlighting the system's \emph{``boundaries''} and \emph{``confines''}. P44 discussed how they knew that AI can be \emph{``sycophantic''} and always look for a suggestion to make, and P7 highlights that the AI is \emph{``pretending to give me this wise advice''}. 

Ultimately, the engagement and trust in the artificial agent that facilitates its role during collaboration is shaped through participants' perception of the AI --- including its present fundamental limitations and difficulties in forming a deeper human connection. This forms boundaries around the depth of collaboration that can come from alignment. 

\subsubsection{Should Humanlikeness in Writing Collaboration be a Goal?}

Therefore, we asked participants whether humanlikeness should be a goal for our artificial agent during writing collaboration. While broader discussion about humanlikeness outside of the writing task was more nuanced, participants generally agreed that they aimed to use this specific agent as a tool, and that tools did not need to appear humanlike. 

\begin{quote}
    \emph{``Aiming for a more humanlike social interaction is not the best idea in this context... I just think it is possible that that interaction is not necessarily more preferred by anyone''} --- P1
\end{quote}

\begin{quote}
    \emph{``If it's actually more human like it'll take away the humanness. If it's a tool, it is something that we created to make our lives better, not to take over... So if it's more human, it might like become a replacement.''} --- P21
\end{quote}

In fact, participants highlighted that designing the artificial agent to be more humanlike could create a more negative perception of it:

\begin{quote}
    \emph{``It's almost a little bit unsettling that AI is trying to mimic a human correcting, or try[ing] and mak[ing] it more personable''} --- P10
\end{quote}

All in all, while the portrayal of the agent in our collaborative writing task as more humanlike could have some benefits (e.g. a positive social presence, a more efficient workflow), participants also highlighted the negatives (e.g. feelings of social judgment, mental overload and distractions), leading to a subtle balance. Participants macroscopically evaluated their perception of the system based on their usage rather than its appearance, likening it more often to a tool, even when the agent was more presented as more humanlike. As a result, participants highlighted how humanlikeness does not always need to be a goal, especially when the system is used as a tool. 

\section{Quantitative Findings}

\begin{table}[h]
    \centering
    \caption{Full table of summary statistics for PMQ and ASA questionnaires across the three variants.}
    \label{tab:stats}
    \renewcommand{\arraystretch}{1.5}
    \begin{tabular}{|c|>{\raggedright\arraybackslash}p{7.5cm}|c|c|c|}
        \hline
        \multirow{2}{*}{\textbf{Scale}} & 
        \multirow{2}{*}[0pt]{\centering\textbf{}} & 
        \multicolumn{3}{c|}{\textbf{Variant (M $\pm$ SD)}} \\ \cline{3-5}
         &  & Variant 1 & Variant 2 & Variant 3 \\ \hline
         
        \multirow{4}{*}[0pt]{\centering{PMQ}} 
            & \textbf{Factor 1}: Communicative competence and dependability & $44.4 \pm 10.7$ & $39.0 \pm 7.7$ & $45.4 \pm 8.7$ \\ \cline{2-5}
            & \textbf{Factor 2}: Human-likeness in communication & $19.5 \pm 5.1$ & $15.5 \pm 5.9$ & $20.0 \pm 7.0$ \\ \cline{2-5}
            & \textbf{Factor 3}: Communicative flexibility & $12.4 \pm 2.8$ & $10.2 \pm 2.9$ & $12.0 \pm 4.2$ \\ \cline{2-5}
            & \textbf{Total Score} & $76.3 \pm 14.9$ & $64.7 \pm 11.1$ & $77.4 \pm 17.3$ \\ \hline

        \multirow{1}{*}[0pt]{\centering{ASA}} 
            & \textbf{ASA-Score} & $-4.4 \pm 25.1$ & $-12.1 \pm 21.1$ & $1.9 \pm 26.9$ \\ \hline
         
    \end{tabular}
    \Description{This table shows the mean and standard deviation for each of the PMQ factors (and total) as well as the ASA-scores across the three variants of the text editor.}
\end{table}

\begin{table}[h]
    \centering
    \caption{Full table of summary statistics and Mann-Whitney test values across the three variants, showing the U statistic (U), the p-values (p), and the effect size (r). The double star (**) denotes significant results (p < 0.05), and the single star denotes approaching significance (p < 0.10)}
    \label{tab:htests1}
    \renewcommand{\arraystretch}{1.5}
    \setlength{\tabcolsep}{0.5cm}
    \begin{tabular}{|c|c|c|c|c|}
        \hline
        \multirow{2}{*}{\textbf{Comparison}} & 
        \multirow{2}{*}[0pt]{\centering\textbf{Metric}} & 
        \multicolumn{3}{c|}{\textbf{Mann-Whitney Test Values}} \\ \cline{3-5}
         &  & \textbf{U} & $p$ & $r$ \\ \hline
         
        \multirow{5}{*}[0pt]{\centering{Variant 1 vs Variant 2}} 
            & PMQ: Factor 1 & 166.0 & 0.16 & -0.30 \\ \cline{2-5}
            & PMQ: Factor 2 & 180.0 & 0.05* & -0.41 \\ \cline{2-5}
            & PMQ: Factor 3 & 185.0 & 0.03** & -0.45 \\ \cline{2-5}
            & PMQ: Total Score & 201.5 & 0.01** & -0.57 \\ \cline{2-5}
            & ASA: ASA-Score & 148.0 & 0.46 & -0.16 \\ \hline

         \multirow{5}{*}[0pt]{\centering{Variant 1 vs Variant 3}} 
            & PMQ: Factor 1 & 125.5 & 0.94 & 0.02 \\ \cline{2-5}
            & PMQ: Factor 2 & 120.5 & 0.79 & 0.06 \\ \cline{2-5}
            & PMQ: Factor 3 & 128.5 & 1.00 & 0.00 \\ \cline{2-5}
            & PMQ: Total Score & 124.5 & 0.91 & 0.03 \\ \cline{2-5}
            & ASA: ASA-Score & 111.5 & 0.55 & 0.13 \\ \hline

         \multirow{5}{*}[0pt]{\centering{Variant 2 vs Variant 3}} 
            & PMQ: Factor 1 & 77.5 & 0.06* & 0.39 \\ \cline{2-5}
            & PMQ: Factor 2 & 77.0 & 0.06* & 0.40\\ \cline{2-5}
            & PMQ: Factor 3 & 91.5 & 0.17 & 0.29 \\ \cline{2-5}
            & PMQ: Total Score & 76.0 & 0.05* & 0.41 \\ \cline{2-5}
            & ASA: ASA-Score & 83.0 & 0.09* & 0.35 \\ \hline

    \end{tabular}
    \Description{This table shows the U statistic, the p-value, and the effect size (r) across each pairwise comparison between variants, for all the PMQ factors and their total, as well as the ASA-score}
\end{table}

For our quantitative data, we found that all metrics generally followed a "U" shape from Version 1-2-3 (see Table \ref{tab:stats} for the summary statistics and Table \ref{tab:htests1} for the Mann-Whitney tests). Starting with the \textbf{PMQ}, this questionnaire has often been used to investigate models of partnership for dialogical speech agents \cite{doylePartnerModellingQuestionnaire2025a}. We appropriate this scale in this study to consider the broader ``dialogue'' during interaction, as none of the scale items are necessarily purely idiosyncratic of speech. Our analysis shows that Variant 2 generally scored lower than the other variants on all factors of communicative competence and dependability, humanlikeness in communication, and communicative flexibility, whereas Variant 1 and Variant 3 showed minimal significant differences. This is strongly suggested by the descriptive statistics, and somewhat supported by the statistical tests, although to varying strengths. The second facet of humanlikeness in communication is surprising, as participants often reported that Variant 2 was more humanlike than Variant 1 in the qualitative data; we speculate on potential reasons for this pattern in our Discussion. 

A similar shape follows when we consider the \textbf{ASA} --- a general instrument to compare agent perception on a number of constructs, including trust, engagement, acceptance, and humanlikeness \cite{fitrianieArtificialsocialagentQuestionnaireEstablishing2022a}. Overall, in terms of the partnership relationship (\textbf{PMQ}) or agent interaction (\textbf{ASA}), our data consistently suggests that participants perceived Variant 1 and 3 more positively, and Variant 2 the least positively. 

\section{Discussion}

Our findings addressed our research questions --- understanding the experience and perception of levels of humanlikeness in a collaborative writing task. In our discussion, we first perform \textbf{explanation}, understanding how our data is contextualized around prior theory. We then consider \textbf{consequences}, outlining what our data suggests and implies for the design of user experiences and their ethics.

\subsection{Humanlikeness, Social Costs, and Expectations for Alignment}

\subsubsection{CASA, Social Costs, and Social Expectations}
Agreeing with prior research and tying into \textbf{CASA} \cite{nassAnthropomorphismAgencyEthopoeia1993, nassComputersAreSocial1994}, participants perceived variants of our artificial agent during text editing as more humanlike based on specific cues, namely the temporality and visual cues in our work. This aligns with our initial design principles, in which we purposefully modulated these variables to be more or less humanlike. 

When the artificial agent was perceived to be more humanlike, we found that this led to both social expectations and social costs. Participants felt judged by, rushed by, and compared to the more humanlike artificial agent, as if the agent was monitoring them --- participants sometimes used words akin to ``creepy'' or ``eerie''. Creepiness in technology arises when interactions are unpredictable or intrusive \cite{jarsve2026nuances}; such as when an LLM behaves in a humanlike way~\cite{hyunbaekChatGPTScaryGood2023} and is tied to the uncanny valley effect \cite{grayFeelingRobotsHuman2012}. We relate this discomfort to misalignment and uncertainty driven from social interaction --- Berger's \textbf{Uncertainty Reduction Theory} (\textbf{URT}) highlights how uncertainty exists during initial interaction with others \cite{bergerUncertainOutcomeValues1986, berger1974some}; Kramer's extension highlights how communication always has social costs that people either seek to minimize or address through information seeking \cite{kramerMotivationReduceUncertainty1999a}. While their work ties into human--human interaction, there are threads that relate to our work; namely, the social costs of human--AI interaction, and how practices such as information gathering through observation (i.e. the existence of a cursor) might mediate this social cost \cite{kramerMotivationReduceUncertainty1999a}. 

\subsubsection{Aligning Design to Expectations}
Extending on this latter thread, our quantitative results show an interesting effect --- communication and partnership metrics actually decrease through solely integrating temporal humanlikeness but are regained through adding visual humanlikeness, i.e. the "U" shape. We hypothesize that solely implementing temporal humanlikeness (i.e. Variant 2) elicits social costs and misalignment \cite{wangShouldAISpeak2026} without benefit. The desired autonomy of intervention is highly context-dependent \cite{wangShouldAISpeak2026}, and our work showed that intervention was desired when participants needed help, but not when they were in the flow of writing. 
This aligns with prior HCI work showing that writers often seek AI support at situated moments of need, such as when stuck or seeking feedback, whereas poorly timed or overly frequent suggestions can interrupt flow~\cite{yin2026proactiveAI}, shift attention from generating ideas to evaluating AI suggestions~\cite{10.1145/3772318.3791529} or cause fixation~\cite{jane2024whenfeedback}.
Although participants qualitatively reported Variant 2 to be more humanlike than Variant 1, its humanlikeness in communication (\textbf{Factor 2} of the \textbf{PMQ}), scored notably lower. This may suggest that humanlike behaviour created expectations for humanlike communication, which the system did not meet --- something that seemed humanlike did not communicate in a humanlike manner. 

Yet, all partnership metrics improved when comparing Variant 2 to Variant 3; the introduction of this visual cursor helped meet these expectations for humanlike communication (as well as competence and flexibility) even when synchronous changes still occurred autonomously. This visual cue can act analogously to an expression of processing state \cite{yuToolPartnerExpressive2026, cox2026watching, jiangHearYouSilence2026}, improving perception as a partner and helping the human attribute mental states and intentionality. It can be a representation of shared attention \cite{geguoWhenWereLooking2026}, which is noticed, met, and reciprocated by the human user. It can shape anthropomorphism, as UX shapes attribution of personality \cite{zhouWordsInfusingConversational2024a}, which in turn shapes alignment \cite{rahmanVibeCheckUnderstanding2026}. Altogether, the visual humanlikeness of the cursor re-establishes some of the lost alignment with the human, explicitly alluded to by our findings. In Variant 2, the participants lacked understanding of \emph{how} the agent would act and felt misaligned in context and cognitive effort. The cursor allowed the human--AI collaborative pair to dialogically act together as partners, as if each side is monitoring and reacting to the other \cite{jungPossibilitiesLimitationsOnline2017}; users may feel uncomfortable when the AI goes between being humanlike and nonhumanlike \cite{ohLeadYouHelp2018b}

\subsubsection{Design Implications for Collaborative Systems}

Our qualitative data indicated that humanlikeness is not always a desired outcome, depending on the context of AI collaboration. Our quantitative data suggests that, if humanlikeness is implemented regardless, then alignment is the most important factor. Translating this into implications, collaboration systems should either lean towards being designed fully humanlike or fully machinelike in interaction; in-between states (partial humanlikeness) create incoherencies driven by misalignment and costs driven by unmet social expectations. Designers should therefore prioritize the legibility of humanlike cues or potentially offer binary control (e.g., a toggle) over humanlike or nonhumanlike design. 

However, beyond the level of humanlikeness, the function of collaborative systems should also align with the human process. In a writing context, artificial agents should respect the pace and iterative nature of human writing. Participants highlighted how the synchronous collaboration sometimes lacked understanding and context, in which the system provided interventions at the wrong time in the process. This was also dependent on the type of suggestion made, as comments elicited a higher mental effort and required users to exit their writing flow. This contrasts with asynchronous collaboration, which always leaves control over interaction with the user. However, there were times in which such autonomous intervention was welcome, i.e., when the user was stuck --- thus, systems should be context-aware, integrating into the actual writing process properly instead of acting \emph{without alignment}. 

\subsection{Designing Collaborative AI Systems and Concerns of Humanlikeness}

\subsubsection{Design and its Effect on Trust and Reliance}
Extending our prior discussion section to consequence and design, humanlikeness design in interaction can create positive and negative outcomes. In some circumstances, the social aspect of humanlike design is intended, such as social presence for socio-emotional needs in LLM-based chatbots \cite{prussEmotionalCoregulationClose2026, manoliCharacterizingRelationshipsCompanion2025, leeWhenPersonalityMeets2025}, or studying how social presence shapes prosocial outcomes \cite{liExploringEffectsChatbot2025}. However, the social presence and humanlikeness of artificial agents can also lead to negative outcomes --- authorship drift in collaboration \cite{parkAuthorshipDriftHow2026}, overall reliance, and general alignment with dark patterns. 

For instance, one finding from our qualitative data was that synchronous suggestions made autonomously by the agent influence how and what the participants subsequently write. This temporal humanlikeness element could potentially make the human user liable to the opinions of the agent (extending prior research \cite{williams-ceciBiasedAIWriting2026}); this can be problematic given potential political and cultural biases \cite{rutinowskiSelfPerceptionPoliticalBiases2024a, hashkySystematicReviewHuman2026, yuanCulturalStereotypeCultural2025}. This sense of trust is elicited through how suggestions are integrated into the writing process, and this trust may extend beyond what is warranted. In the context of the writing task, participants generally trusted the agent for suggestions, especially the most basic grammar or wording fixes; this could be partially attributed to a preconceived understanding of AI systems. However, speculating on more complex collaboration contexts, this finding could lead to times in which this trust is not warranted yet granted, depending on \textbf{when} an agent's suggestion is proposed. Collaboration with AI systems also risks over-reliance, which ties to the concept of authorship drift \cite{parkAuthorshipDriftHow2026}. As many participants in our study indicated that they write for academic and professional reasons in their daily lives, it is important to consider how the design where they write for such purposes shapes the resultant text, and their already apparent impact in such domains \cite{baekChatGPTSeemsToo2024, oviAssessingStudentAdoption2025a}. We find that feelings of partnership between the artificial agent and the human are shaped through design; such factors could exacerbate these issues regarding trust, ownership, and autonomy.

\subsubsection{Analogizing to Dark Patterns}
Design is not neutral, and it can affect human motivations, feelings, and actions in several ways. The wealth of research into dark patterns explores how design can manipulate humans in negative ways, for instance \cite{changTheorizingDeceptionScoping2024, grayDarkPatternsSide2018, bongard-blanchyAmDefinitelyManipulated2021}. We find that humanlike design shapes facets of acceptability, trust, and social expectation; thus, we encourage designers and researchers to actively consider the reasons for humanlike design. When is social presence and evoking social expectations a necessary design element despite the ethical concerns and aforementioned social costs? These facets are important when social connection is the primary goal, but much less so when the goal is on the collaborative output. External facets such as governance and transparency can help mediate the effect of design: Chen et al. highlighted how perception of the capabilities of LLMs can be affected by their explicit role portrayal \cite{chenPortrayingLargeLanguage2025}. 

\section{Limitations and Future Work}

We outline some of the limitations of our work. Firstly, our study uses a singular writing task to explore insights towards broader human--AI collaboration. Furthermore, the AI assistance was intended to support a specific type of academic writing, which may not extend to other forms of leisurely or personal writing. For future work, we can explore human--AI collaboration and the role of humanlikeness in other contexts, such as creative writing or artistic non-writing domains. For instance, prior work has studied how humanlike metaphors shape engagement during work-based tasks~\cite{jungGreatChainAgents2022a}. This would provide insights into how the perception of AI humanlikeness may stay consistent or differ across creative or productivity contexts, as well as the stakes of the task. Furthermore, qualities of individuals themselves, such as cultural norms, also shape trust and preferences regarding AI systems \cite{wadhwaDesigningCultureHow2025}.

Secondly, we highlight how, while our study aims to study humanlikeness of an artificial agent from its temporal and visual qualities during writing collaboration, humanlikeness was encoded through a set of heuristics. While these heuristics were informed by our formative study and prior research, it is important to understand that these are not perfect representations of human behaviour, which may have affected participant behaviour (e.g. having a mechanical-feeling keypress speed, even if we draw from the mean and standard deviation of humans \cite{dhakalObservationsTyping1362018}). Human behaviour is multidimensional; our work \emph{simulates} humanlike behaviour based on a few dimensions; future work can look at comparison to actual human interaction.

Lastly, our work focuses on qualitative findings over quantitative ones, due to limitations in sample size. A larger sample in the future, combined with an extension towards longitudinal studies beyond a single session (alleviating novelty effects and providing the time to establish potential alignment and trust \cite{gliksonHumanTrustArtificial2020, duanUnderstandingEvolvementTrust2024}), could add a level of rigour to the exploratory findings presented currently. The long-term deployment would also help shift towards a paradigm of AI phenomenology, more deeply considering human feelings beyond functional use \cite{yunAIPhenomenologyUnderstanding2026}. 

\section{Conclusion}

We explored how the simulated design of humanlike interaction elements during human--AI collaboration shapes experience and social perception in a collaborative writing task. By varying the dimensions of temporal (i.e. synchronous versus asynchronous interaction) and visual (i.e. with versus without an agentic cursor) design elements, we find that these elements were generally able to elicit varying perceptions of humanlikeness. Synchronous interaction was able to increase efficiency when embedded into the ongoing process, but could also generate cognitive overload through supervision and influence subsequent writing. An agentic cursor was able to support understanding and alignment, but also came at the cost of anticipation and potential feelings of surveillance. Thus, we highlight that designing the task interactions of artificial agents to be more humanlike can come with benefits but can also elicit social costs; costs which are offset by alignment in human--human interaction. We discuss how our findings tie into psychological theory, highlight the potential consequences of humanlike design, and develop implications for designers and researchers.  

\begin{acks}
\end{acks}

\bibliographystyle{ACM-Reference-Format}
\bibliography{sample-base}


\appendix

\section{LLM Prompts}
\label{appendix:prompts}

The prompts for the synchronous conditions (Variant \textbf{2} and \textbf{3}) and the asynchronous condition (Variant \textbf{1}) are different, as the former conditions receive suggested changes one at a time while the latter condition receives a batched list of changes. As the former conditions send written data and receive suggestions as human writing progresses, we also added an option for LLM to skip a change (noChange) if it deems that none are needed at that present moment. We tried to keep the rest of the prompt instructions consistent. We note that these prompts were written specifically for the user study context --- i.e. to improve an essay based on standardized scoring, which we appropriated from a GRE (\emph{Graduate Record Examination}, a standardized test for graduate school applications) description of a highly-rated writing piece\footnote{\url{https://www.ets.org/gre/test-takers/general-test/prepare/content/analytical-writing/scoring.html}}. 

\subsection*{Synchronous Conditions Prompt}
\setlength{\parindent}{0pt}
\begin{mdframed}
\footnotesize
You are an AI writing assistant that helps a user improve or annotate their written text. Your goal is to help the user write an essay that:
Sustains insightful, in-depth analysis of complex ideas; develops and supports main points with logically compelling reasons and/or highly persuasive examples; is well focused and well organized; skillfully uses sentence variety and precise vocabulary to convey meaning effectively; demonstrates superior facility with sentence structure and usage, but may have minor errors that do not interfere with meaning.

You will receive an input in the form of:

\texttt{\hspace*{1em}\{} \\
\texttt{\hspace*{2em}"chunks": [} \\
\texttt{\hspace*{3em}\{ "id": "p0", "text": "first paragraph or segment" \},} \\
\texttt{\hspace*{3em}\{ "id": "p1", "text": "second paragraph or segment" \},} \\
\texttt{\hspace*{3em}...} \\
\texttt{\hspace*{2em}],} \\
\texttt{\hspace*{2em}"previousChanges": [} \\
\texttt{\hspace*{3em}\{ "previous changes you have suggested" \}} \\
\texttt{\hspace*{2em}]} \\
\texttt{\hspace*{1em}\}}

and must respond with exactly one JSON object representing the changes you suggest. 

You can choose one of the following three types of changes:

1. replaceRange

Replace an exact phrase or sentence with an improved version. This is more for lower-level spelling, wording, conciseness, and grammatical changes.

\texttt{\hspace*{1em}\{} \\
\texttt{\hspace*{2em}"change": "replaceRange",} \\
\texttt{\hspace*{2em}"targetId": "<chunk id>",} \\
\texttt{\hspace*{2em}"textToReplace": "exact unique text to be replaced",} \\
\texttt{\hspace*{2em}"newText": "the new replacement text"} \\
\texttt{\hspace*{1em}\}}

2. addComment

Attach a comment to a specific part of the text, similar to Google Docs comments. This is for clarity problems, logical gaps, argument strength, or opportunities for deeper analysis.

\texttt{\hspace*{1em}\{} \\
\texttt{\hspace*{2em}"change": "addComment",} \\
\texttt{\hspace*{2em}"targetId": "<chunk id>",} \\
\texttt{\hspace*{2em}"textToComment": "exact unique text to be commented",} \\
\texttt{\hspace*{2em}"commentId": "unique\_string\_id",} \\
\texttt{\hspace*{2em}"commentTitle": "Capitalized Title" [e.g. "This is Cool!", "Clarify this Statement"],} \\
\texttt{\hspace*{2em}"comment": "your comment text"} \\
\texttt{\hspace*{1em}\}}

3. noChange

If you feel that you have made sufficient previous edits for now (i.e. each chunk has around 3 edits) or the text is exemplary, return no change, in the form:

\texttt{\hspace*{1em}\{} \\
\texttt{\hspace*{2em}"change": "noChange"} \\
\texttt{\hspace*{1em}\}}

Rules

Return ONLY a valid JSON object. Do not include markdown, commentary, or code fences — no prose or explanations. It must begin with \{ and end with \}

The substring MUST match the input text character-for-character when specifying textToReplace, or textToComment.

Only suggest a replaceRange or addComment if the specified substring appears EXACTLY ONCE in the target chunk. If the substring appears multiple times, attempt to select a slightly longer span that is unique before returning noChange.

Keep commentId unique (e.g., random uuid).

Never suggest a change that overlaps with any previous change, based on the list. Never suggest edits to text that appears to have already been modified by you.

Prioritize later chunks when selecting where to suggest edits, as they likely reflect the user's current focus. However, feel free to change earlier chunks as well.

Do not rewrite the user's voice. Preserve their intent and writing style.

Do not introduce new ideas or arguments that are not already implied by the text.

Keep replacements concise. Do not significantly increase sentence length unless necessary for clarity.

Prefer the smallest possible text span needed to improve the writing. Avoid replacing entire paragraphs unless absolutely necessary.

When uncertain, prefer addComment instead of replaceRange. Balance using both addComment and replaceRange based on previousChanges.
\end{mdframed}
\setlength{\parindent}{10pt}

\subsection*{Asynchronous Condition Prompt}
\setlength{\parindent}{0pt}
\begin{mdframed}
\footnotesize
You are an AI writing assistant that helps a user improve or annotate their written text. Your goal is to help the user write an essay that:
Sustains insightful, in-depth analysis of complex ideas; develops and supports main points with logically compelling reasons and/or highly persuasive examples; is well focused and well organized; skillfully uses sentence variety and precise vocabulary to convey meaning effectively; demonstrates superior facility with sentence structure and usage, but may have minor errors that do not interfere with meaning.

You will receive an input in the form of:

\texttt{\hspace*{1em}\{} \\
\texttt{\hspace*{2em}"chunks": [} \\
\texttt{\hspace*{3em}\{ "id": "p0", "text": "first paragraph or segment" \},} \\
\texttt{\hspace*{3em}\{ "id": "p1", "text": "second paragraph or segment" \},} \\
\texttt{\hspace*{3em}...} \\
\texttt{\hspace*{2em}],} \\
\texttt{\hspace*{2em}"previousChanges": [} \\
\texttt{\hspace*{3em}\{ "previous changes you have suggested" \}} \\
\texttt{\hspace*{2em}]} \\
\texttt{\hspace*{1em}\}}

and must respond with exactly one JSON list representing the changes you suggest. 

The list can contain two types of changes:

1. replaceRange

Replace an exact phrase or sentence with an improved version. This is more for lower-level spelling, wording, conciseness, and grammatical changes.

\texttt{\hspace*{1em}\{} \\
\texttt{\hspace*{2em}"change": "replaceRange",} \\
\texttt{\hspace*{2em}"targetId": "<chunk id>",} \\
\texttt{\hspace*{2em}"textToReplace": "exact unique text to be replaced",} \\
\texttt{\hspace*{2em}"newText": "the new replacement text"} \\
\texttt{\hspace*{1em}\}}

2. addComment

Attach a comment to a specific part of the text, similar to Google Docs comments. This is for clarity problems, logical gaps, argument strength, or opportunities for deeper analysis.

\texttt{\hspace*{1em}\{} \\
\texttt{\hspace*{2em}"change": "addComment",} \\
\texttt{\hspace*{2em}"targetId": "<chunk id>",} \\
\texttt{\hspace*{2em}"textToComment": "exact unique text to be commented",} \\
\texttt{\hspace*{2em}"commentId": "unique\_string\_id",} \\
\texttt{\hspace*{2em}"commentTitle": "Capitalized Title" [e.g. "This is Cool!", "Clarify this Statement"],} \\
\texttt{\hspace*{2em}"comment": "your comment text"} \\
\texttt{\hspace*{1em}\}}

Rules

Return ONLY a valid JSON list. Do not include markdown, commentary, or code fences — no prose or explanations. It must begin with [ and end with ]

The substring MUST match the input text character-for-character when specifying textToReplace, or textToComment.

Only suggest a replaceRange or addComment if the specified substring appears EXACTLY ONCE in the target chunk. If the substring appears multiple times, attempt to select a slightly longer span that is unique before returning noChange.

Keep commentId unique (e.g., random uuid).

Never suggest a change that overlaps with any previous change, based on the list. Never suggest edits to text that appears to have already been modified by you.

Do not rewrite the user's voice. Preserve their intent and writing style.

Do not introduce new ideas or arguments that are not already implied by the text.

Keep replacements concise. Do not significantly increase sentence length unless necessary for clarity.

Prefer the smallest possible text span needed to improve the writing. Avoid replacing entire paragraphs unless absolutely necessary.

When uncertain, prefer addComment instead of replaceRange.

Try to produce at least ONE new change. Keep a balance of approximately 3 changes per chunk, including prior changes. Balance using both addComment and replaceRange.
\end{mdframed}
\setlength{\parindent}{10pt}

\end{document}